\documentclass[prc,twocolumn,showpacs,preprintnumbers,amsmath,amssymb,superscriptaddress,floatfix,nofootinbib]{revtex4}
\usepackage{graphicx}%
\begin{document}
%%%%%%%%%%%%%%%%%%%%%%%%%%%%%%%%%%%%%%%%%%%%%%%%%%%%%%%%%%%%%%%%%%%%%%
\title{Faddeev fixed center approximation to the $N\bar{K}K$ system and
the signature of a $N^*(1920)(1/2^+)$ state}
\author{Ju-Jun Xie}
\email{xiejujun@ific.uv.es} \affiliation{Instituto de F\'\i sica
Corpuscular (IFIC), Centro Mixto CSIC-Universidad de Valencia,
Institutos de Investigaci\'on de Paterna, Aptd. 22085, E-46071
Valencia, Spain} \affiliation{Department of Physics, Zhengzhou
University, Zhengzhou, Henan 450001, China}
\author{A. Mart\'inez Torres}
\email{amartine@yukawa.kyoto-u.ac.jp} \affiliation{Yukawa Institute
for Theoretical Physics, Kyoto University, Kyoto 606-8502, Japan}
\author{E. Oset}
\email{oset@ific.uv.es}
\affiliation{Instituto de F\'\i sica
Corpuscular (IFIC), Centro Mixto CSIC-Universidad de Valencia,
Institutos de Investigaci\'on de Paterna, Aptd. 22085, E-46071
Valencia, Spain} \affiliation{Departamento de F\'{\i}sica
Te\'{o}rica, Universidad de Valencia, Valencia, Spain.}

\preprint{YITP-11-15}
\begin{abstract}

We perform a calculation for the three body $N \bar{K} K$ scattering
amplitude by using the fixed center approximation to the Faddeev
equations, taking the interaction between $N$ and $\bar{K}$, $N$ and
$K$, and $\bar{K}$ and $K$ from the chiral unitary approach. The
resonant structures show up in the modulus squared of the three body
scattering amplitude and suggest that a $N\bar{K}K$ hadron state can
be formed. Our results are in agreement with others obtained in
previous theoretical works, which claim a new $N^*$ resonance around
$1920$ MeV with spin-parity $J^P=1/2^+$. The existence of these
previous works allows us to test the accuracy of the fixed center
approximation in the present problem and sets the grounds for
possible application in similar problems, as an explorative tool to
determine bound or quasibound three hadron systems.

\end{abstract}
\pacs{14.20.Gk.; 21.45.-v.}
\maketitle

\section{Introduction}

The study of hadron structure and the spectrum of hadron resonances
is one of the most important issues and is attracting much
attention. Recently, the topic of meson-baryon states, with mesons
and baryons governed by strong interactions, is well developed by
the combination of the chiral Lagrangians with nonperturbative
unitary techniques in coupled channels, which has been a very
fruitful scheme to study the nature of many hadron resonances. The
analysis of meson-baryon and meson-meson scattering amplitudes shows
poles in the second Riemanm sheet, which are identified with
existing hadron resonances or new ones. In this way the low-lying
$1/2^-$ resonances are
generated~\cite{kaiser1535,osetkbarN,reciokbarN,hyodokbarN,inouepiN,juankbarN,hyodoweisekbarN,jido1405},
including the $N^*(1535)$~\cite{kaiser1535} and two $\Lambda(1405)$
states~\cite{jido1405}. More recently, the low-lying $1/2^+$ states
have also been generated from two mesons and a baryon such as the
$N^*(1710)$ resonance and others~\cite{alberto17101,alberto17102}.

Within the chiral unitary approach, the $f_0(980)$ and $a_0(980)$
scalar mesons are dynamically generated from the interaction of
$\bar{K}K$, $\pi \pi$, and $\eta \pi$ treated as coupled channels in
$I=0$ and $I=1$,
respectively~\cite{ollerkbark1,ollerkbark2,ollerkbark3,nicolakbark,pelaezkbark,kaiserkbark,markushinkbark}.
Both couple strongly to the $\bar{K}K$ channel. In this sense the
$f_0(980)$ and $a_0(980)$ scalar mesons can be explained as $\bar{K}
K$ bound states.

The effective interactions of $\bar{K}N$ have been studied in a
unitary chiral
theory~\cite{osetkbarN,reciokbarN,hyodokbarN,juankbarN,hyodoweisekbarN,jido1405,ollerchiral1,borasoychiral1,ollerchiral2,borasoychiral2}
and in a phenomenological
way~\cite{Mares:2004pf,Shevchenko:2006xy,Shevchenko:2007zz,Cieply:2009ea}.
Within the unitary chiral theory, two $\Lambda(1405)$ states are
dynamically generated, as we mentioned above, the higher mass one
corresponding to basically a $\bar{K}N$ bound state and the lower
mass one looking like a $\pi \Sigma$ resonance. In the
phenomenological approaches, the strong attraction of $\bar{K}N$ in
the $I=0$ sector provides $\Lambda(1405)$ as a bound state of
$\bar{K}N$.

Recently, there is growing evidence that some existing hadronic
states can be interpreted in terms of bound states or resonances of
three hadrons. Some new states of this nature have also been
reported. For example, it has been found that the $Y(4660)$
resonance reported in $e^+e^- \to \gamma_{\text{ISR}}\pi^+\pi^-
\psi'$ can be interpreted as a $\psi' f_0(980)$ bound
state~\cite{guo4660}. In Ref.~\cite{albertopikN}, the Faddeev
equations for the $\pi \bar{K} N$ system were solved by using the
chiral unitary approach and coupled channels, and several $\Sigma^*$
and $\Lambda^*$ resonances with spin-parity $J^P=1/2^+$ in the
energy region $1500-1800$ MeV were dynamically generated. The
$X(2175)$ (now $\phi(2170)$) state, claimed by different
experimental groups~\cite{ex21751,ex21752,ex21753} in the $\phi
f_0(980)$ invariant mass spectrum, has been interpreted as a $\phi
\bar{K}K$ resonance with $\bar{K}K$ forming the $f_0(980)$ scalar
meson~\cite{alberto2175,oller2175,coito2175}.

For the $N \bar{K} K$ system, having strong attraction in the
$\bar{K} N$ and $\bar{K} K$ subsystems, it is naturally expected
that the three hadrons $N \bar{K} K$ form a hadron state. Indeed,
this state has been studied in Ref.~\cite{jido1920} with
nonrelativistic three-body variational calculations by using
effective $\bar{K} N$, $\bar{K} K$, and $K N$ interactions and in
Refs.~\cite{alberto19201,alberto19202,alberto19203} by solving the
Faddeev equations in a coupled channel approach. They all found a
bound state of the $N \bar{K}K$ system with total isospin $I=1/2$
and spin-parity $J^P=1/2^+$.

In the present work, we reinvestigate the three-body $N \bar{K} K$
system by considering the interaction of the three components among
themselves, keeping in mind the expected strong correlations of the
$\bar{K} K$ and $\bar{K} N$ system to make the $f_0(980)(a_0(980))$
and $\Lambda(1405)$ respectively. In terms of two-body $N\bar{K}$,
$NK$ and $K\bar{K}$, $KN$ scattering amplitudes, we solve the
Faddeev equations by using the Fixed Center Approximation(FCA). The
main purpose of the present work is to test the validity of the FCA
to study systems of three hadrons in conditions similar to the
present one. The FCA has been employed before, in particular in the
study of the $\bar{K} d$ interaction at low
energies~\cite{Chand:1962ec,Barrett:1999cw,Deloff:1999gc,Kamalov:2000iy}.
Comparison of full Faddeev equations and the FCA can be extracted
from Refs.~\cite{Toker:1981zh,Gal:2006cw}, where it is shown that
the FCA is a good approximation for this problem at the level of a
few percent. This approach was also used in Ref.~\cite{fcarhorho} to
describe the $f_2(1270)$, $\rho_3(1690)$, $f_4(2050)$,
$\rho_5(2350)$ and $f_6(2510)$ resonances as multi-$\rho(770)$
states, and in Ref.~\cite{YamagataSekihara:2010qk} to study the
$K^*_2(1430)$, $K_3^*(1780)$, $K^*_4(2045)$, $K^*_5(2380)$, and a
not yet discovered $K^*_6$ resonances as $K^*-$multi$-\rho$ states.
Very recently, we gave an interesting explanation of the
$\Delta_{5/2^+}$ puzzle by using this approach in the $\pi \Delta
\rho$ system~\cite{Xie:2011uw}.

One basic feature of the FCA is that one has a cluster of two
particles and one allows the multiple scattering of the third
particle with this cluster, which is supposed not to be changed by
the interaction of the third particle. Intuitively one can think
that this would mostly happen when the particles in the cluster are
more massive than the third one. One might also think that the
approximation could be better if the third particle had small
energy, such that it can not perturb too much the cluster of the two
particles. Whatever it is, one does not know the accuracy of the FCA
until a comparison is made with the more elaborate Faddeev
equations. On the other hand, technically, the FCA is much easier
than the Faddeev equations, which require lengthly calculations and
also approximations, to the point that some approximations done to
make the Faddeev equations feasible might introduce larger
uncertainties in some problem than those introduced by the FCA. With
this perspective it is important to be able to quantify
uncertainties of the FCA in certain circumstances such that it can
be used as a prospective tool to investigate possible structures of
three hadrons, leading to bound or quasibound states, in different
systems and similar circumstances. This is the main purpose of the
present work. We shall take advantage that the present problem has
been solved with two independent methods, full Faddeev
equations~\cite{alberto19201,alberto19202,alberto19203} and
variational approach~\cite{jido1920}, such that comparison with the
result of these works can give us a feeling of the accuracy of the
FCA.

In next section, we present the FCA formalism and ingredients to
analyze the $N\bar{K}K$ system. In Section III, our results and
discussions are presented. Finally, conclusions are given in section
IV.

\section{Formalism and ingredients}

We are going to use the FCA of the Faddeev equations in order to
obtain the scattering amplitude of the three body $N \bar{K} K$
system. We consider $\bar{K} K$ as a bound state of $f_0/a_0(980)$
scalar meson in one case, and $\bar{K} N$ as a bound state of
$\Lambda(1405)$ state in the other, which allows us to use the FCA
to solve the Faddeev equations. The analysis of the
$N-(\bar{K}K)_{f_0/a_0(980)}$ and $K-(\bar{K}N)_{\Lambda(1405)}$
scattering amplitudes will allow us to study dynamically generated
resonances.

The important ingredients in the calculation of the total scattering
amplitude for the $N \bar{K} K$ system using the FCA are the
two-body $\bar{K} N$, $K N$ and $\bar{K}K$ unitarized $s-$wave
interactions from the chiral unitary approach. Since this work has
been reported on many occasions, we direct the reader for details to
Refs.~\cite{osetkbarN,ollerkbark1,Oset:2001cn}.

In the FCA approach we need as input the wave function of a two
particle cluster and the scattering amplitude of the third particle
with those of the clusters. The wave functions, and the
corresponding form factors, are taken from
Refs.~\cite{danielprd81,YamagataSekihara:2010pj}, where a quantum
mechanical study of the coupled channels interaction is done in a
way that gives the same scattering amplitudes as those obtained in
the field theoretical treatment of the chiral unitary approach.

\subsection{Form factors for the $f_0/a_0(980)$ and the $\Lambda(1405)$}

One of the ingredients in the calculation is the form factor for the
assumed two body cluster, the $f_0(980)$ and $a_0(980)$ in one case,
or the $\Lambda(1405)$ in the other. Following the approach of
Ref.~\cite{danielprd81,YamagataSekihara:2010pj}, we can easily get
the expression for the form factor $F_{B^*}(q)$ for the bound state
$B^*$ of a pair of particles~\footnote{For the sake of brevity and
to avoid repeating similar equations for the two different
configurations, the field normalization factors in our approach
include a factor $2m_N$ for nucleon, but since we normalize
$F_{B^*}$ to 1 at $q=0$, Eqs.~(\ref{ff}) and (\ref{ffN}) are
general.},
\begin{eqnarray}
&& F_{B^*}(q) = \frac{1}{{\cal N}} \int_{|\vec{p}|<\Lambda,
|\vec{p}-\vec{q}|<\Lambda} d^3\vec{p} \frac{1}{2\omega_1(\vec{p})}
\frac{1}{2\omega_2(\vec{p})} \times \nonumber \\
&& \frac{1}{M-\omega_1(\vec{p})-\omega_2(\vec{p})}
\frac{1}{2\omega_1(\vec{p}-\vec{q})}
\frac{1}{2\omega_2(\vec{p}-\vec{q})} \times \nonumber \\
&& \frac{1}{M-\omega_1(\vec{p}-\vec{q})-\omega_2(\vec{p}-\vec{q})},
\label{ff}
\end{eqnarray}
where the normalization factor ${\cal N}$ is
\begin{eqnarray}
{\cal N} &=& \int_{|\vec{p}|<\Lambda} d^3\vec{p} \left(
\frac{1}{2\omega_1(\vec{p})}
\frac{1}{2\omega_2(\vec{p})} \right)^2 \times \nonumber \\
&& \frac{1}{(M -\omega_1(\vec{p})-\omega_2 (\vec{p}))^2},
 \label{ffN}
\end{eqnarray}
where $M$ is the mass of the bound state, which is $980$ MeV for the
$f_0$ or $a_0$ scalar meson and 1420 MeV for the $\Lambda(1405)$
state in the present computation. The approach requires the
knowledge of $\Lambda$, the cut off that is needed in
Refs.~\cite{danielprd81,YamagataSekihara:2010pj}, to regularize the
loop functions in the chiral unitary approach.

\begin{figure}[htbp]
\begin{center}
\includegraphics[scale=0.45] {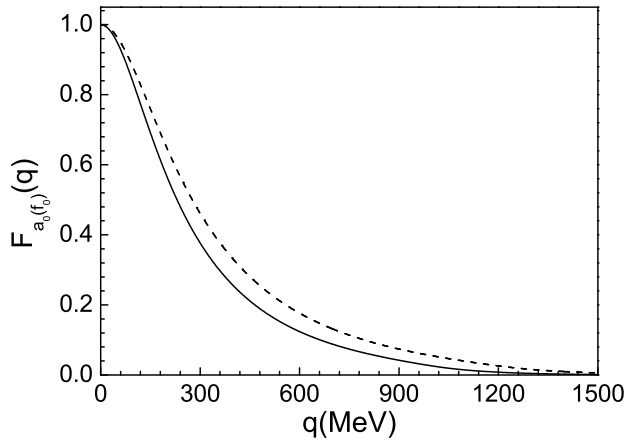}%
\caption{Form factor for the $f_0(980)(a_0(980))$. The solid line
stands for the original result, and the dashed line corresponds to
the cluster with a radius decreased by $20\%$.} \label{Fig:ff980}
\end{center}
\end{figure}

\begin{figure}[htbp]
\begin{center}
\includegraphics[scale=0.45] {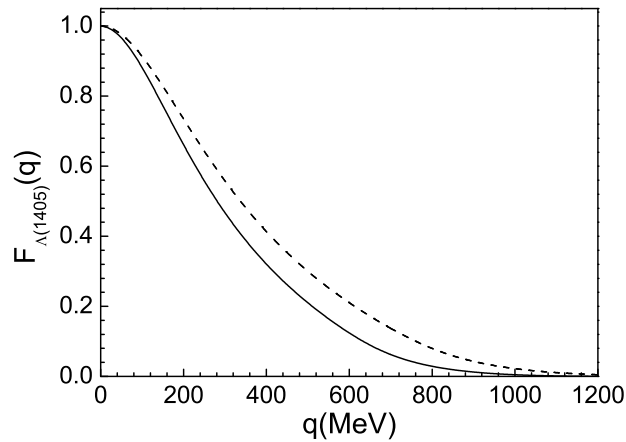}%
\caption{As in Fig.~\ref{Fig:ff980} but for the $\Lambda(1405)$.}
\label{Fig:ff1405}
\end{center}
\end{figure}

In Figs.~\ref{Fig:ff980}, and~\ref{Fig:ff1405}, we show the
respective form factors for the $f_0(a_0(980))$ and $\Lambda(1405)$
as a function of $q$. The condition $|\vec{p}-\vec{q}|<\Lambda$
implies that the form factor is exactly zero for $q>2\Lambda$. In
the present work, we use $\Lambda=1000$ MeV and $630$ MeV for the
$f_0/a_0(980)$ and the $\Lambda(1405)$,
respectively~\cite{osetkbarN,ollerkbark1}.

In the FCA, we keep the wave function of the cluster unchanged by
the presence of the third particle. In order to estimate
uncertainties of the FCA due to this "frozen" condition we admit
that the wave function of the cluster could be modified by the
presence of the third particle. Since the interaction is attractive
in the present case it should induce a reduction of the size of the
cluster~\cite{Dote:2008hw}. However, unlike in
Ref.~\cite{Dote:2008hw} where the size of the $NN$ cluster is mostly
determined by the long range of $\pi$ exchange, and its size can be
sizeable reduced, in the present case both the $\bar{K}K$ and
$\bar{K}N$ interaction are governed by the exchange of vector mesons
(implicit in the chiral Lagrangians and explicitly shown in the
local hidden gauge formalism~\cite{Oset:2009vf}), thus the size is
determined by the constraints of the uncertainty principle, since
the range of the interaction is zero. In this case the presence of
the third particle cannot reduce the size of the cluster much. In
order to quantify uncertainties of the FCA, we perform calculations
for the case where the cluster radius is decreased by $20\%$.
Technically, this is accomplished by increasing $20\%$ the masses of
the particles involved in the cluster in Eqs.~(\ref{ff}) and
(\ref{ffN}).

\subsection{Faddeev FCA equations}

\begin{figure}[htbp]
\includegraphics[scale=0.5]{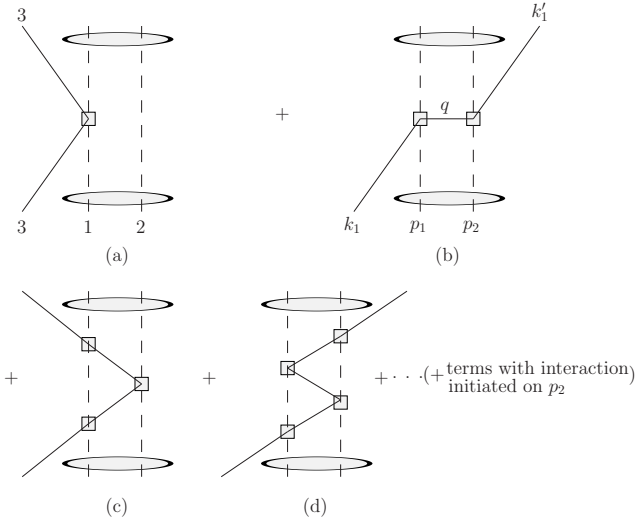}
\caption{Diagramatic representation of the fixed center
approximation to the Faddeev equations. Diagrams (a) and (b)
represent the first contributions to the Faddeev equations from
single scattering and double scattering respectively. Diagrams (c)
and (d) represent iterations of the interaction.} \label{diagram}
\end{figure}

The FCA to the Faddeev equations are depicted diagrammatically in
Fig.~\ref{diagram}. The external particle 3 interacts successively
with the particle 1 and particle 2 which form the bound states. For
the case of $N-(\bar{K}K)_{f_0/a_0(980)}$ configuration, since the
masses of $f_0(980)$ and $a_0(980)$ are equal, the $Nf_0 \to Na_0$
transition should be important to the total three body scattering
amplitude, and we consider this transition in our calculation. Then
the FCA equations for the $N-(\bar{K}K)_{f_0/a_0(980)}$
configuration are written in terms of four partition functions
($T_1$, $T_2$, $\bar{T_1}'$, and $\bar{T_2}'$ for $Nf_0 \to Nf_0$,
and $\widetilde{T_1}$, $\widetilde{T_2}$, $\bar{T_1}$, and
$\bar{T_2}$ for $Na_0 \to Na_0$), which read,
\begin{eqnarray}
&& T_1 = t_1+t_1G_0T_2 + \bar{t_1}G_0\bar{T_2}', \label{Tf01} \\
&& T_2 = t_2+t_2G_0T_1 + \bar{t_2}G_0\bar{T_1}', \label{Tf02} \\
&& \bar{T_1}' = \bar{t_1} + \bar{t_1}G_0T_2 + \widetilde{t_1}G_0\bar{T_2}', \label{Ta0f01} \\
&& \bar{T_2}' = \bar{t_2} + \bar{t_2}G_0T_1 + \widetilde{t_2}G_0\bar{T_1}', \label{Ta0f02} \\
&& T_{Nf_0 \to Nf_0} = T_1+T_2,        \label{TNf0} \\
&& \widetilde{T_1} = \widetilde{t_1} + \widetilde{t_1}G_0\widetilde{T_2} + \bar{t_1}G_0\bar{T_2}, \label{Ta01} \\
&& \widetilde{T_2} = \widetilde{t_2} + \widetilde{t_2}G_0\widetilde{T_1} + \bar{t_2}G_0\bar{T_1}, \label{Ta02} \\
&& \bar{T_1} = \bar{t_1} + \bar{t_1}G_0\widetilde{T_2} + t_1G_0\bar{T_2}, \label{Tf0a01} \\
&& \bar{T_2} = \bar{t_2} + \bar{t_2}G_0\widetilde{T_1} + t_2G_0\bar{T_1}, \label{Tf0a02} \\
&& T_{Na_0 \to Na_0} = \widetilde{T_1} + \widetilde{T_2},
\label{TNa0}
\end{eqnarray}
where $T_{Nf_0 \to Nf_0}$ and $T_{Na_0 \to Na_0}$ are the total
three-body scattering amplitudes, $T_i$ and $\widetilde{T_i}$
($i=1,2$) account for all the diagrams starting from the interaction
of the external particle with particle $i$ of the compound system
for $Nf_0 \to Nf_0$ and $Na_0 \to Na_0$, respectively, while
$\bar{T_i}$ and $\bar{T_i}'$ account for the intermediate virtual
transition contributions for $Nf_0 \to Na_0$ and $Na_0 \to Nf_0$,
respectively. Hence, $t_i$, $\widetilde{t_i}$ and $\bar{t_i}$
represent the corresponding two body (particle 3 and particle 1 or
particle 3 and particle 2) unitarized scattering amplitudes. In the
above equations, $G_0$ is the loop function for the particle 3
propagating inside the compound system which will be discussed later
on.

For the case of the $K-(\bar{K}N)_{\Lambda(1405)}$ configuration,
since we do not expect other $s-$wave meson-baryon channels to play
any relevant role in the present calculation, the FCA equations are
written in terms of only two partition functions $T_1$ and $T_2$,
which sum up to the total three body scattering amplitude. In this
case the equations are easy and we just have
\begin{eqnarray}
&&T_1 = t_1+t_1G_0T_2, \label{T14051} \\
&&T_2 = t_2+t_2G_0T_1, \label{T14052} \\
&& T_{K\Lambda(1405) \to K\Lambda(1405)} = T_1+T_2. \label{TK1405}
\end{eqnarray}

\subsection{Single-scattering contribution}

The amplitude corresponding to the single-scattering contribution
(Fig.~\ref{diagram} (a) $+$ term with interaction initiated on
$p_2$) comes from the $t_1,$ $\widetilde{t_1},$ $\bar{t_1}$ and
$t_2,$ $\widetilde{t_2},$ $\bar{t_2}$, which are the appropriate
combination of the two body (particle 3 and particle 1 or particle 3
and particle 2) unitarized scattering amplitudes. For example, let
us consider a cluster of $\bar{K} K$ in $I=0$ ($f_0(980)$), the
constituents of which we call $1$ and $2$, and the external nucleon
$N$ will be numbered $3$. The $\bar{K} K$ isospin state is written
as
\begin{eqnarray}
|\bar{K} K>_{I=0} &=& \sqrt{\frac{1}{2}}|(\frac{1}{2},
-\frac{1}{2})> - \sqrt{\frac{1}{2}}|(-\frac{1}{2}, \frac{1}{2})> ,
\end{eqnarray}
where the kets in the last member indicate the $I_z$ components of
the particles $1$ and $2$, $|(I_z^{(1)}, I_z^{(2)})>$.

The scattering amplitude $<N\bar{K}K|t|N\bar{K}K>$ for the single
scattering contribution can be easily obtained in terms of the two
body amplitudes $t_{31}$ and $t_{32}$ derived in
Refs.~\cite{osetkbarN,reciokbarN,hyodokbarN,juankbarN,hyodoweisekbarN,jido1405,ollerchiral1,borasoychiral1,ollerchiral2,borasoychiral2}.

Here we write explicitly the case of $Nf_0 \to Nf_0$,
\begin{widetext}
\begin{eqnarray}
& & <N\bar{K}K|t|N\bar{K}K> = (<(\bar{K} K)|_{I_z=0} \bigotimes
<N|_{I_z=1/2}) (t_{31}+t_{32}) (|(\bar{K} K)>_{I_z= 0 } \bigotimes
|N>_{I_z=1/2} ) \nonumber \\
 & = & ( \sqrt{\frac{1}{2}} (<(\frac{1}{2},
-\frac{1}{2})| - <(-\frac{1}{2}, \frac{1}{2})|) \bigotimes
<N|_{I_z=1/2})(t_{31}+t_{32})( \sqrt{\frac{1}{2}}
(|(\frac{1}{2}, -\frac{1}{2})> - |(-\frac{1}{2}, \frac{1}{2})>)  \bigotimes |N>_{I_z=1/2} ) \nonumber \\
&=& < - \sqrt{\frac{1}{2}}((1 1),-\frac{1}{2}) + \frac{1}{2}(((1 0)
+ (0 0)),\frac{1}{2})| t_{31} | - \sqrt{\frac{1}{2}}((1
1),-\frac{1}{2}) + \frac{1}{2}(((1 0) +
(0 0)),\frac{1}{2}) > + \nonumber \\
&& < \sqrt{\frac{1}{2}}((1 1),-\frac{1}{2}) - \frac{1}{2}(((1 0) +
(0 0)),\frac{1}{2})| t_{32}| \sqrt{\frac{1}{2}}((1 1),-\frac{1}{2})
- \frac{1}{2}(((1 0) + (0 0)),\frac{1}{2}) >,
\end{eqnarray}
\end{widetext}
where the notation followed in the last term for the states is
$((I_{\bar{K} N} I_{\bar{K} N}^z),I_K^z)$ for $t_{31}$, while
$((I_{K N} I_{K N}^z),I_{\bar{K}}^z)$ for $t_{32}$. This leads to
the following amplitude for the single scattering contribution,
\begin{eqnarray}
t_1 &=& \frac{1}{4} t_{N \bar{K}}^{I=0} + \frac{3}{4} t_{N
\bar{K}}^{I=1} \equiv \frac{1}{4} t_{N \bar{K}}^{0} + \frac{3}{4}
t_{N \bar{K}}^{1}, \\
t_2 &=& \frac{1}{4} t_{N K}^{I=0} + \frac{3}{4} t_{N K}^{I=1} \equiv
\frac{1}{4} t_{N K}^{0} + \frac{3}{4} t_{N K}^{1}.
\end{eqnarray}

Proceeding in a similar way, we can get all the amplitudes for the
single scattering contribution in the present calculation which are
shown in Table~\ref{tab:2bodyamp} for the cases of
$N-(\bar{K}K)_{f_0/a_0(980)}$ and $K-(\bar{K}N)_{\Lambda(1405)}$
configurations with total isospin $I_{N\bar{K}K} =1/2$ .

\begin{table}[htbp]
\caption{Unitarized two-body scattering amplitudes for the single
scattering contribution for the cases of
$N-(\bar{K}K)_{f_0/a_0(980)}$ and $K-(\bar{K}N)_{\Lambda(1405)}$
configurations with $I_{N\bar{K}K} =1/2$. \label{tab:2bodyamp} }
\begin{center}
\begin{tabular}{|c|c|c|}
  & $t_1 ~~(\widetilde{t_1} ~~\text{or}~~ \bar{t_1})$  & $t_2 ~~(\widetilde{t_2} ~~\text{or}~~ \bar{t_2})$ \\
\vspace*{-0.3cm} && \\
 \hline
\vspace*{-0.3cm} && \\
 $Nf_0 \to Nf_0$ & $\frac{1}{4}t_{N \bar{K}}^0 + \frac{3}{4}t_{N \bar{K}}^1$
& $\frac{1}{4}t_{N K}^0 + \frac{3}{4}t_{N K}^1$ \\
\vspace*{-0.3cm} && \\
\hline \vspace*{-0.3cm} && \\
 $Na_0 \to Na_0$  & $\frac{3}{4}t_{N \bar{K}}^0 + \frac{1}{4}t_{N
 \bar{K}}^1 $ & $\frac{3}{4}t_{N K}^0 + \frac{1}{4}t_{N K}^1 $ \\
\vspace*{-0.3cm} && \\
\hline \vspace*{-0.3cm} && \\
$Nf_0 \to Na_0$  & $\frac{\sqrt{3}}{4}t_{N \bar{K}}^1 - \frac{\sqrt{3}}{4}t_{N \bar{K}}^0$ &
 $ - \frac{\sqrt{3}}{4}t_{N K}^1 + \frac{\sqrt{3}}{4}t_{N K}^0$ \\
\vspace*{-0.3cm} && \\
\hline \vspace*{-0.3cm} && \\
$K\Lambda(1405) \to K\Lambda(1405)$ & $\frac{3}{4}t_{K\bar{K}}^1+\frac{1}{4}t_{K\bar{K}}^0$ & $\frac{3}{4}t_{KN}^1+\frac{1}{4}t_{KN}^0$ \\

\end{tabular}
\end{center}
\end{table}

It is worth noting that the argument of the total scattering
amplitudes $T_{Nf_0 \to Nf_0}$, $T_{Na_0 \to Na_0}$ and
$T_{K\Lambda(1405) \to K\Lambda(1405)}$ are functions of the total
invariant mass $s$, while the argument in $t_1$ $(\widetilde{t_1}
~~\text{or}~~ \bar{t_1})$ is $s_1'$ and in $t_2$ $(\widetilde{t_2}
~~\text{or}~~ \bar{t_2})$ is $s_2'$, where $s_1'$ and $s_2'$ are the
invariant masses of the external particle 3 with momentum $k_1$ and
particle 1 (2) insider the bound state ($f_0(980)$, $a_0(980)$ or
$\Lambda(1405)$) with momentum $p_1$($p_2$), which are given by
\begin{eqnarray}
s_1' &=&  m^2_3 + m^2_1 + \nonumber \\
&& \frac{(M^2+m^2_1-m^2_2)(s-m^2_3-M^2)}{2M^2},  \label{s1prime} \\
s_2' &=&  m^2_3 + m^2_2 +  \nonumber \\
&& \frac{(M^2+m^2_2-m^2_1)(s-m^2_3-M^2)}{2M^2}.  \label{s2prime}
\end{eqnarray}

Following the approach developed in Ref.~\cite{fcarhorho}, (see also
section $4$ of Ref.~\cite{YamagataSekihara:2010pj} for details on
the form factors) we can easily write down the $S-$matrix for the
single scattering term in the Mandl-Shaw normalization~\cite{mandl},
which we follow, as

\begin{eqnarray}
S^{(1)} &=& S_1^{(1)} + S_2^{(1)} \nonumber \\
&=&  ( (-i t_1 F_{B^*}(\frac{m_2(\vec{k_1}-\vec{k_1')}}{m_1+m_2}))
\frac{BF_1}{{\cal V}^2}  \frac{1}{\sqrt{2\omega_1}}
\frac{1}{\sqrt{2\omega_1'}} +  \nonumber  \\
&& (-i t_2 F_{B^*}(\frac{m_1(\vec{k_1}-\vec{k_1')}}{m_1+m_2}))
\frac{BF_2}{{\cal V}^2}  \frac{1}{\sqrt{2\omega_2}}
\frac{1}{\sqrt{2\omega_2'}} ) \times \nonumber  \\
&& \frac{1}{\sqrt{2\omega_3}} \frac{1}{\sqrt{2\omega_3'}}  (2\pi)^4
\delta^4(k_1+K-k_1'-K'), \label{s1}
\end{eqnarray}
where $\mathcal{V}$ stands for the volume of a box where we
normalize to unity our plane wave states, and $BF_1$ and $BF_2$ are
the field normalization baryon factors, $BF_1=BF_2=2m_N$ for
$N-(\bar{K}K)_{f_0(a_0(980))}$ configuration, while for the case of
$K-(\bar{K}N)_{\Lambda(1405)}$, they are $BF_1=1$, $BF_2=2m_N$. With
this normalization, $t_1$ and $t_2$ have the standard form as
evaluated in the chiral unitary approach. The symbols $k_1$, $k_1'$,
$K$, and $K'$ stand for the four momenta of the initial, final
particle 3 and initial, final bound state $B^*$. In Eq.~(\ref{s1}),
$F_{B^{\ast}}(\frac{m_i(\vec{k_{1}
}-\vec{k_{1}^{\prime})}}{m_1+m_2})$ is the form factor of the bound
state $B^*$. This form factor was taken to be unity neglecting the
$\vec{k},\vec{k^{\prime}}$ momentum in Ref.~\cite{fcarhorho} since
only states below threshold were considered. To consider states
above threshold, we project the form factor into $s$-wave, the only
partial wave that we consider. Thus

\begin{align}
F_{B^{*}}(\frac{m_i(\vec{k_{1}}-\vec{k^{\prime}_{1})}}{m_1+m_2})
\Rightarrow FFS_i(s) =\frac {1}{2} \int_{-1}^{1} F_{B^{*}}(k_i)
d(cos\theta),
\end{align}
with
\begin{align}
k_i= \frac{m_i}{m_1+m_2} k \sqrt{2(1-cos\theta)}, \quad
\left\{\begin{array}{l} i=1 ~ {\rm for} ~ t_2 ~(\widetilde{t_2}
~\text{or}~ \bar{t_2}), \cr i=2 ~ {\rm for} ~ t_1 ~(\widetilde{t_1}
~\text{or}~ \bar{t_1}).
\end{array}\right. \nonumber
\end{align}
and
\begin{align}
k =\frac{\sqrt{(s-(m_1+m_2 + m_3)^{2})(s-(m_1+m_2 - m_3)^2)}
}{2\sqrt{s}}, \label{mk}
\end{align}
is the modulus of the momentum of the particle 3 in the center of
mass frame of $N\bar{K}K$ system when the total rest energy
$\sqrt{s}$ is above the threshold of the three particles. Below
threshold, the wave functions still provide a finite real momentum
$k$, but are small enough, such that we can take $k=0$ in this case.

\begin{figure}[ptbh]
\begin{center}
\includegraphics[scale=0.45] {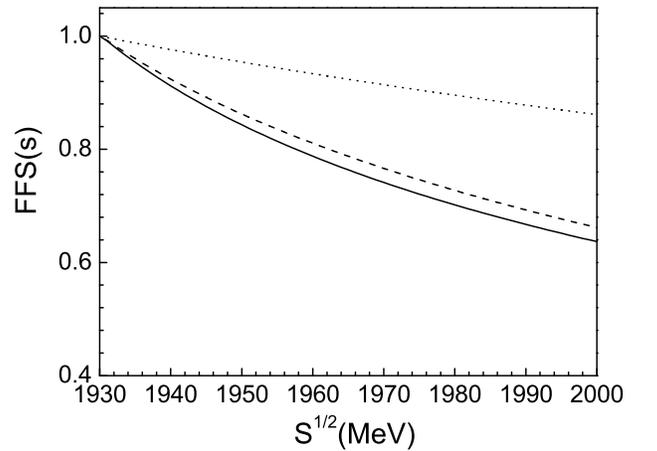}
\end{center}
\caption{Form factor $FFS(s)$ for the $Nf_0(a_0(980))$ and
$K\Lambda(1405)$ scattering.}%
\label{Fig:ffs}%
\end{figure}

In Fig.~\ref{Fig:ffs}, we show the projection over the s-wave of the
form factor $FFS_1(s)$ (equal to $FFS_2(s)$) (solid line) for the
case of $Nf_0(a_0(980))$ scattering and $FFS_1(s)$ (dashed line),
$FFS_2(s)$ (dotted line) for the case of $K\Lambda(1405)$
scattering. As one can see, in all cases, the form factor has a
smooth behavior.

\subsection{Double-scattering contribution}

Next, we are going to evaluate the amplitude of the
double-scattering contribution (Fig.~\ref{diagram} (b) + term with
interaction initiated on $p_2$) in the same way as in case of
multi-rho meson interaction in Ref.~\cite{fcarhorho}. The expression
for the $S-$matrix for the double scattering ($S_2^{(2)} =
S_1^{(2)}$), in the Mandl-Shaw normalization~\cite{mandl}, is,

\begin{eqnarray}
S_1^{(2)} &=& (-i t_1 t_2) (2\pi)^4 \delta^4(k_1+K -k_1'-K') \times
\nonumber \\ && \frac{BF_1BF_2}{{\cal V}^2}
\frac{1}{\sqrt{2\omega_1}} \frac{1}{\sqrt{2\omega_1'}}
\frac{1}{\sqrt{2\omega_2}} \frac{1}{\sqrt{2\omega_2'}}
\frac{1}{\sqrt{2\omega_3}}
\frac{1}{\sqrt{2\omega_3'}} \nonumber \\
&& \times \int \frac{d^3\vec{q}}{(2\pi)^3} F_{B^*}(q) D(q^{2}), \label{s2}%
\end{eqnarray}
where $BF_1=BF_2=\sqrt{2m_N}$ for $Nf_0(a_0(980))$ scattering, or
$BF_1=1$ and $BF_2=2m_N$ for the $K\Lambda(1405)$ scattering. The
propagator $D(q^{2})$ is
\begin{eqnarray}
D(q^2)=\frac{m_N}{E_N(\vec{q}~^{2})}\frac{1}{q^0-E_N(\vec{q}~^{2}) +
i\epsilon}
\end{eqnarray}
for the case of $Nf_0(a_0(980))$ scattering, and
\begin{eqnarray}
D(q^2)=\frac{1}{q^{0^{2} }-\vec{q}~^{2}-m^{2}_K + i\epsilon}
\end{eqnarray}
for the case of $K\Lambda(1405)$ scattering.

One must now take into account that the field normalization factor
for the $Nf_0(a_0(980))$ or $K\Lambda(1405)$ scattering amplitude
are neither those of Eq.~(\ref{s1}) or Eq.~(\ref{s2}). But

\begin{eqnarray}
S &=& -i T \frac{ BF_1BF_2}{{\cal V}^2}
\frac{1}{\sqrt{2\omega_{B^*}}} \frac{1}{\sqrt{2\omega_{B^*}'}}
\frac{1}{\sqrt{2\omega_3}} \frac{1}{\sqrt{2\omega_3'}} \times \nonumber \\
&& (2\pi)^4 \delta^4(k_1+K-k_1'-K'). \label{s}
\end{eqnarray}
where $BF_1=2m_N$, $BF_2=1$ for $Nf_0(a_0(980))$ scattering and
$BF_1=1$, $BF_2=2M_{\Lambda(1405)}$ for the $K\Lambda(1405)$
scattering.

By comparing Eq.~(\ref{s}) with Eq.~(\ref{s1}) for the
single-scattering and Eq.~(\ref{s2}) for the double-scattering, we
see we have to give a weight to $t_1$ and $t_2$ such that
Eqs.~(\ref{s1}) and (\ref{s2}) get the weight factors that appear in
the general formula of Eq.~(\ref{s}). Taking this into account, the
terms in the multiple scattering of the FCA can be summed up with
Eqs.~(\ref{Tf01}$-$\ref{TNa0}) and
Eqs.~(\ref{T14051}$-$\ref{T14052}) by means of

\begin{eqnarray}
t_1 \rightarrow && t_1
\sqrt{\frac{2\omega_{f_0(a_0(980))}}{2\omega_{\bar{K}}}}
\sqrt{\frac{2\omega_{f_0(a_0(980))}'}{2\omega_{\bar{K}}'}}, \nonumber \\
t_2 \rightarrow && t_2
\sqrt{\frac{2\omega_{f_0(a_0(980))}}{2\omega_{K}}}
\sqrt{\frac{2\omega_{f_0(a_0(980))}'}{2\omega_{K}'}}, \nonumber \\
G_0(s) &=& \frac{1}{\sqrt{2\omega_{f_0(a_0(980))}
2\omega_{f_0(a_0(980))}'}} \times \nonumber \\
&& \int \frac{d^3\vec{q}}{(2\pi)^3} F_{f_0(a_0(980))}(q)
\frac{m_N}{E_N(\vec{q}~^{2})}\frac{1}{q^0-E_N(\vec{q}~^{2}) +
i\epsilon}, \nonumber
\end{eqnarray}
for the case of $N-(\bar{K}K)_{f_0(a_0(980))}$ configuration,

\begin{eqnarray}
t_1 \rightarrow && t_1 \frac{1}{\sqrt{2\omega_{\bar{K}}}}
\frac{1}{\sqrt{2\omega_{\bar{K}}'}}
\sqrt{\frac{E_{\Lambda(1405)}}{{M_{\Lambda(1405)}}}} \sqrt{\frac{E_{\Lambda(1405)}'}{{M_{\Lambda(1405)}}}}, \nonumber \\
t_2 \rightarrow && t_2 \sqrt{\frac{m_N}{E_N}}
\sqrt{\frac{m_N}{E_N'}}
\sqrt{\frac{E_{\Lambda(1405)}}{{M_{\Lambda(1405)}}}} \sqrt{\frac{E_{\Lambda(1405)}'}{{M_{\Lambda(1405)}}}}, \nonumber \\
G_0(s) &=& \sqrt{\frac{M_{\Lambda(1405)}}{E_{\Lambda(1405)}}} \sqrt{\frac{M_{\Lambda(1405)}}{E_{\Lambda(1405)}'}} \times \nonumber \\
&& \int \frac{d^3\vec{q}}{(2\pi)^3} F_{\Lambda(1405)}(q)
\frac{1}{q^{0^{2} }-\vec{q}~^{2}-m^{2}_K + i\epsilon}, \nonumber
\end{eqnarray}
for the case of $K-(\bar{K}N)_{\Lambda(1405)}$ configuration.

\begin{figure}[htbp]
\begin{center}
\includegraphics[scale=0.45]{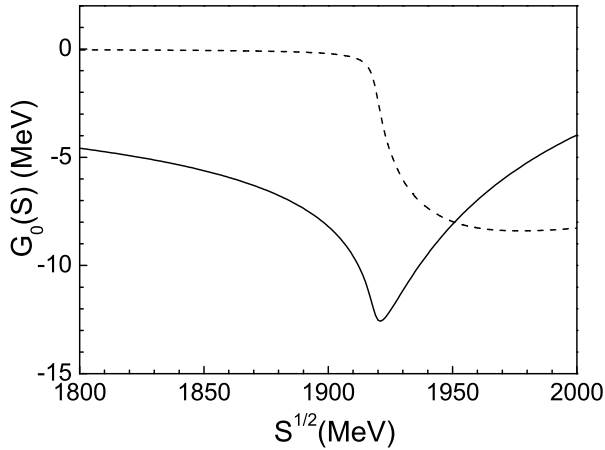}
\vspace{0.cm} \caption{Real (solid line) and imaginary (dashed line)
parts of the $G_0$ as a function of the total energy for the case of
$Nf_0(a_0(980))$ scattering.}
\label{Fig:gzero980} %
\end{center}
\end{figure}

\begin{figure}[htbp]
\begin{center}
\includegraphics[scale=0.45]{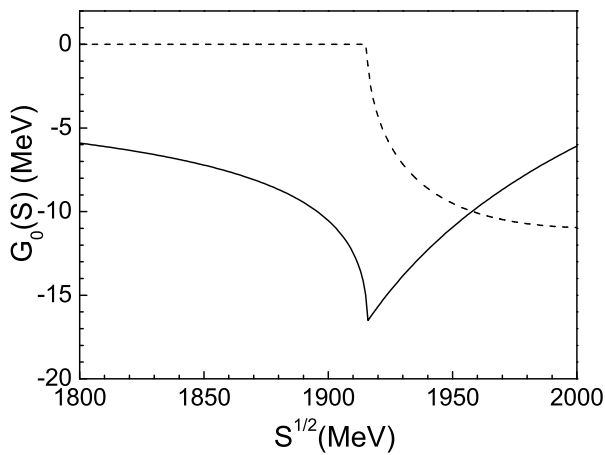}%
\caption{Real (solid line) and imaginary (dashed line) parts of the
$G_0$ as a function of the total energy for the case of
$K\Lambda(1405)$ scattering.}
\label{Fig:gzero1405} %
\end{center}
\end{figure}

In Figs.~\ref{Fig:gzero980} and \ref{Fig:gzero1405}, we show the
real and imaginary parts of the loop function $G_0$ as a function of
the invariant mass of the $N \bar{K} K$ system for the two cases,
where the value of $q^0 = (s + m^{2}_3 - M^{2})/(2\sqrt{s})$ is
given by the energy of particle 3 in the center of mass frame of the
particle 3 and the bound state $B^*$.

\section{Results and discussions}

In this section we show the results obtained for the scattering
amplitude of the $N \bar{K} K$ system with total isospin $I=1/2$ and
spin-parity $J^P=1/2^+$. We evaluate the scattering amplitude $T$
and associate the peaks in the modulus squared $|T|^2$ to
resonances.

\begin{figure}[htbp]
\begin{center}
\includegraphics[scale=0.45]{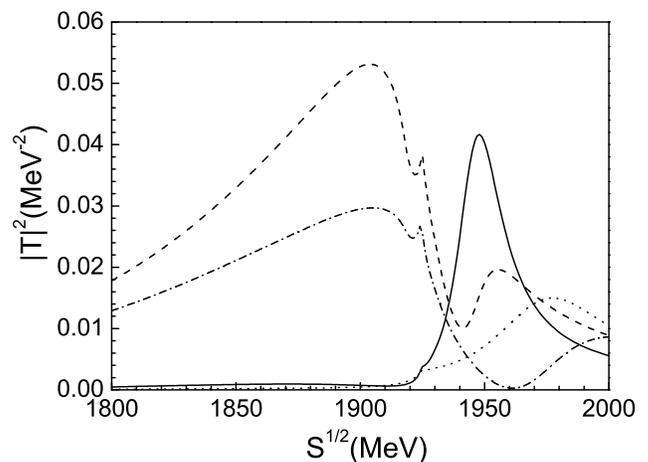}%
\vspace{-0.5cm} \caption{Modulus squared of the $Nf_0(a_0(980))$
scattering amplitude in $I_{\text{total}}=1/2$. The solid line and
dashed line stand for the $Nf_0 \to Nf_0$ scattering and the $Na_0
\to Na_0$ scattering with the original form factors, respectively.
The dotted ($Nf_0 \to Nf_0$) and dash-dotted ($Na_0 \to Na_0$) stand
for the results obtained with the modified form factor of the
cluster $\bar{K}K$ with a radius decreased by $20\%$.}
\label{Fig:1half980}%
\end{center}
\end{figure}

In Fig.~\ref{Fig:1half980}, we show the modulus squared $|T|^2$ for
the $Nf_0 \to Nf_0$ and $Na_0 \to Na_0$ scattering. The picture
shows a clear peak for the $Na_0 \to Na_0$ case around $1915$ MeV,
and a width of about $80$ MeV~\footnote{Although we mentioned that
the wave functions provide finite momenta for the $N$, formally we
could extrapolate the form factor below threshold, making the $N$
momentum purely imaginary. If we do that, the peak shifts to $1925$
MeV, and the width is around $30$ MeV. We mention these results as
an indication of possible uncertainties below threshold.}. There is
no peak at this energy for $Nf_0 \to Nf_0$, but, a narrower peak
appears around $1950$ MeV, which also appears with smaller strength
for the case of $Na_0 \to Na_0$ scattering.

The peak appearing for $Na_0 \to Na_0$ could be associated with the
one obtained in Ref.~\cite{jido1920}. It should be noted that taking
into account the transition from $Nf_0 \to Na_0$ is important.
Although the two structures in Fig.~\ref{Fig:1half980} also appear
when omitting the transition, its consideration leads to a much
wider distribution around the peak in the case of the $Na_0 \to
Na_0$ amplitude. We can also mention that in the case of $Na_0 \to
Na_0$ with total isospin $I=3/2$, we also obtain a peak around
$1960$ MeV. However, the strength is very small compared to the case
of $I=1/2$, too small to have a significant impact on a realistic
amplitude when other components would necessarily mix in the wave
functions. Yet, there are still more considerations to make. Indeed,
we have assumed that the structure of this three body system
corresponds basically to a cluster of $\bar{K}K$, making either
$f_0(980)$ or $a_0(980)$, and a nucleon. It could be that the
physical state would choose better another configuration where the
$\bar{K}N$ cluster makes the $\Lambda(1405)$. Indeed, the study of
Ref.~\cite{alberto19203} showed that there was a correlation of
$\bar{K}N$ around the $\Lambda(1405)$, which was also mentioned in
Ref.~\cite{jido1920}. In what follows, we pay attention to this
other configuration.

\begin{figure}[htbp]
\begin{center}
\includegraphics[scale=0.45]{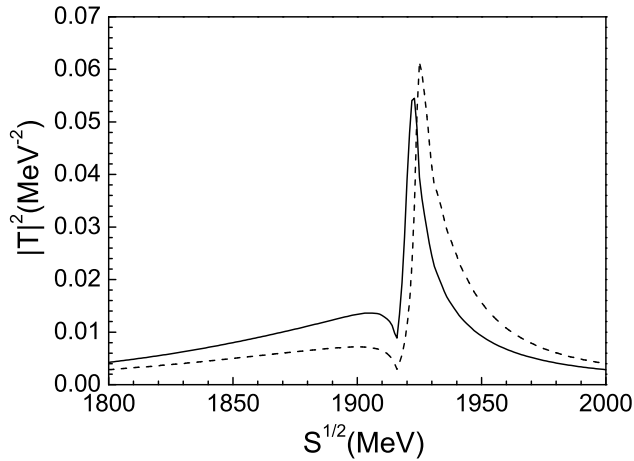}  %
\vspace{-0.5cm} \caption{Modulus squared of the $K\Lambda(1405)$
scattering amplitude. The solid line stands for the results obtained
by using the original form factors, while the dashed line stands for
the results obtained with the modified form factor corresponding to
the cluster of $\bar{K}N$ with a radius decreased by $20\%$.}%
\label{Fig:1half1405}%
\end{center}
\end{figure}

In Fig.~\ref{Fig:1half1405}, we show the results of $|T|^2$ for the
case of $K\Lambda(1405) \to K\Lambda(1405)$. Here we see a clear
peak around $1925$ MeV. The strength of $|T|^2$ at the peak is
similar to that of Fig.~\ref{Fig:1half980} for the $Na_0 \to Na_0$
scattering. A proper comparison requires to reconsider the
normalization factors in the $S$ matrix in the case of
$K\Lambda(1405) \to K\Lambda(1405)$ versus those of $Na_0 \to Na_0$,
as seen in Eqs.~(\ref{s}). It is then clear that the proper
comparison is $T_{Na_0 \to Na_0}$ versus
$\frac{M_{a_0(980)}}{m_K}T_{K\Lambda(1405) \to K\Lambda(1405)}$.
Taking this into account we find that $|\frac{M_{a_0(980)}}{m_K}
T_{K\Lambda(1405) \to K\Lambda(1405)}|^2 \simeq 4 |T_{Na_0 \to
Na_0}|^2$. It is thus clear that the preferred configuration is
$K\Lambda(1405)$. However, the $K$ will keep interacting with the
$\bar{K}$ or the $N$ and sometimes can also make an $f_0(980)$ or
$a_0(980)$, as one can see in the Table~\ref{tab:2bodyamp}, which
means that both the $a_0(980)$ and $\Lambda(1405)$ or $f_0(980)$ and
$\Lambda(1405)$ configuration would be present as found in
Refs.~\cite{alberto19201,alberto19202,alberto19203}.

Note that the peak of Fig.~\ref{Fig:1half1405} is relatively narrow.
Yet, once the $K\Lambda(1405)$ configuration would mix with the
$Na_0(980)$ configuration of Fig.~\ref{Fig:1half980}, the mixture
would produce a wider shape. The important thing is that both
configurations peak around the same energy and hence the peak around
this energy of any mixture of the states is guaranteed.

The large dominance of the $K\Lambda(1405)$ component over the
$Nf_0(a_0(980))$ serves to put the peaks with moderate strength
around $1950$ MeV seen in Fig.~\ref{Fig:1half980} in a proper
context, indicating that the effect of this configuration in that
energy region can be diluted when other large components of the wave
functions are considered, such that we should not expect that these
peaks would have a much repercussion in any physical observable.

As we mentioned earlier, we would like to test the stability of the
results by releasing the assumption of the FCA, ie, that the cluster
of two particles on which the third one scatters is "frozen" in the
process of multiple scattering. For this purpose we repeat the
calculation using the reduced size of the $\bar{K}K$ or $\bar{K}N$
clusters, decreasing their radii in $20\%$. The results can be seen
in Figs.~\ref{Fig:1half980} and \ref{Fig:1half1405}. The basic
features of Fig.~\ref{Fig:1half980} remain, the peak corresponding
to the $Na_0$ component being visible but having become less sharp,
and the same happens with the $Nf_0$ component. Yet, the most
important thing, as shown in Fig.~\ref{Fig:1half1405}, is that the
$K\Lambda(1405)$ component, which is by far the largest as we argued
before, is practically unchanged, showing a neat peak around $1925$
MeV.

In summary, what is solid from our analysis is that one finds a
clear peak in the scattering amplitude for the $N\bar{K}K$ system,
indicating that we have a resonant state made of these components,
built up mostly around the $K\Lambda(1405)$ configuration and where
the $\bar{K}K$ subsystem has an important overlap with the
$f_0(980)$ and $a_0(980)$ resonances.

The stability of the main component, and the agreement of the
results with the more sophisticated ones of
Refs.~\cite{jido1920,alberto19201,alberto19202,alberto19203}, shows
that the FCA is a reliable tool for this problem and this gives
support to its use in similar problems and conditions in other three
hadron systems.

\section{Conclusions}

We have conducted a study of the $N\bar{K}K$ system assuming that
there is a primary clustering of the particles as $Nf_0(980)$,
$Na_0(980)$, or $K\Lambda(1405)$. By using the FCA to the Faddeev
equations we have observed in both $Na_0(980)$ and $K\Lambda(1405)$
configurations there is a clear peak around $1920$ MeV indicating
the formation of a resonant $N\bar{K}K$ state around this energy. We
also found that the $K\Lambda(1405)$ configuration is the dominant
one, where the $\bar{K}K$ subsystem can still couple to the
$f_0(980)$ and $a_0(980)$ resonances. The results obtained here are
in agreement with those obtained in Ref.~\cite{jido1920} using a
variational procedure and in
Refs.~\cite{alberto19201,alberto19202,alberto19203} using full
Faddeev couple channels calculations, which support the existence of
a $N^*$ state with spin-parity $J^P=1/2^+$ around $1920$ MeV. In
Ref.~\cite{alberto19201} some arguments were given in favor of the
peak seen in $\gamma p \to K^+
\Lambda$~\cite{ex19101,ex19102,ex19103} around this energy being due
to this resonance (other suggestions that do not exclude the one of
Ref.~\cite{alberto19201} are given in
Refs.~\cite{Nikonov:2007br,Schumacher:2010qx,delaPuente:2008bw}).

Since the independent works of Ref.~\cite{jido1920} and
Refs.~\cite{alberto19202,alberto19203}, together with that of
Ref.~\cite{alberto19201}, already give strong support for the
$N^*(1920)(1/2^+)$ state, the main value of the present paper is not
to provide extra support for this state but to test the reliability
of the FCA to deal with the $N\bar{K}K$ system. We found the main
results of the FCA to be rather stable and this gives us confidence
that one can use this, technically easier, tool to study similar
systems of three hadrons where the conditions are similar. The fact
that the energies of the particles are small and one is close to
threshold of the three particle system should be viewed as one
important condition for a reliable FCA result. Should the scattering
particle have a relatively large energy, one could no longer invoke
that the cluster of the other two particles remain unchanged. This
is intuitive, but it has been made quantitative in
Ref.~\cite{MartinezTorres:2010ax} in the system made by $\phi
\bar{K}K$ in connection with the $\phi(2170)$ resonance. Works like
the present one and that of Ref.~\cite{MartinezTorres:2010ax} are
setting the grounds for the applicability of the FCA, which should
help make an exploration of likely bound three hadron systems in a
span of time substantive shorter than with the lengthy and
technically complicated full Faddeev calculations.

\section*{Acknowledgments}

This work is partly supported by DGICYT Contract No. FIS2006-03438,
the Generalitat Valenciana in the project PROMETEO, the Spanish
Consolider Ingenio 2010 Program CPAN (CSD2007-00042) and the EU
Integrated Infrastructure Initiative Hadron Physics Project under
contract RII3-CT-2004-506078. Ju-Jun Xie acknowledges Ministerio de
Educaci\'{o}n Grant SAB2009-0116. The work of A.~M.~T.~is supported
by the Grant-in-Aid for the Global COE Program \textquotedblleft The
Next Generation of Physics, Spun from Universality and Emergence"
from the Ministry of Education, Culture, Sports, Science and
Technology (MEXT) of Japan.

\end{document}